\documentclass[twocolumn]{aastex62}
\usepackage{graphicx}
\usepackage{bm}
\usepackage{amssymb}
\usepackage{amsmath}
\usepackage{units}
\usepackage{array}

\submitjournal{\apjl}
\received{2020 October 16}
\revised{2021 January 12}
\accepted{2021 January 15}
\published{2021 March 3}
\reportnum{arXiv:2010.09707}
\reportnum{doi:10.3847/2041-8213/abdd1c}

\shorttitle{Eccentric accreting binaries}
\shortauthors{Zrake et al.}

\begin{document}

\title{Equilibrium eccentricity of accreting binaries}

\correspondingauthor{Jonathan Zrake}
\email{jzrake@clemson.edu}

\author[0000-0002-1895-6516]{Jonathan Zrake}
\affiliation{Department of Physics and Astronomy, Clemson University, Clemson, SC 29634, USA}

\author[0000-0002-3820-2404]{Christopher Tiede}
\affiliation{Center for Cosmology and Particle Physics, Physics Department, New York University, New York, NY 10003, USA}

\author[0000-0002-0106-9013]{Andrew MacFadyen}
\affiliation{Center for Cosmology and Particle Physics, Physics Department, New York University, New York, NY 10003, USA}

\author[0000-0003-3633-5403]{Zolt\'an Haiman}
\affiliation{Department of Astronomy, Columbia University, New York, NY 10027, USA}

\begin{abstract}

Using high-resolution hydrodynamics simulations, we show that equal-mass binaries accreting from a circumbinary disk evolve toward an orbital eccentricity of $e \simeq 0.45$, unless they are initialized on a nearly circular orbit with $e \lesssim 0.08$, in which case they further circularize. The implied bi-modal eccentricity distribution resembles that seen in post-AGB stellar binaries. Large accretion spikes around periapse impart a tell-tale, quasi-periodic, bursty signature on the light curves of eccentric binaries. We predict that intermediate-mass and massive black hole binaries at $z \lesssim 10$ entering the \emph{LISA} band will have measurable eccentricities in the range $e \simeq 10^{-3} - 10^{-2}$, if they have experienced a gas-driven phase. On the other hand, GW190521 would have entered the LIGO/Virgo band with undetectable eccentricity $\sim 10^{-6}$ if it had been driven into the gravitational wave regime by a gas disk.

\end{abstract}


%
\section{Introduction} \label{sec:intro}

Gas accretion onto orbiting binaries is of general importance to the understanding of a range of astrophysical systems. Examples include massive black hole binaries \citep{Begelman1980}, binary protostars \citep[e.g.][and references therein]{Alves2019}, post-AGB stellar binaries \citep[e.g.][]{Mathieu1991}, and stellar-mass objects embedded in accretion disks in galactic nuclei
\citep{Baruteau+2011}. Each of these binaries may be surrounded by, and accrete from, a circumbinary gas disk at some stage of its life.

The gas flow established from the binary-disk interaction exerts gravitational forces on the binary, and directly transfers mass and momentum to it, leading to evolution of the system's orbital elements over time. Epochs of gas accretion could thus be responsible for the present-day eccentricities of many binary systems. The accretion dynamics might also be reflected in the light curves of all types of accreting binaries, and could be sensitive to the system's orbital parameters. It is therefore valuable to understand how binary eccentricities evolve in response to accretion, and in turn how their accretion signals behave as a function of eccentricity.

It has been suggested as early as \cite{Artymowicz1992} that the eccentricities of accreting comparable-mass binary stars might be driven up as high as $e \gtrsim 0.7$, based on linear theory \citep{Goldreich1980} and clues from early smooth-particle hydrodynamics (SPH) simulations \citep{Artymowicz1991}. \cite{Armitage2005} and \cite{Cuadra2009} have both reported simulations of binaries evolving to $e \gtrsim 0.3$ (see their Figs. 2 and 8 respectively). More recent SPH simulations \citep{Roedig2011} showed unequal mass binaries with $M_2/M_1 = 1/3$ evolving toward an eccentricity between $0.6$ and $0.8$. Moving-mesh hydrodynamic simulations \citep{Munoz2019} of equal-mass binaries have exhibited eccentricity growth at $e=0.1$, and suppression at $e=0.5$ and $0.6$.

In this {\it Letter} we report simulations showing that circumbinary accretion tends to evolve equal-mass binaries toward a stable eccentricity $e_{\rm eq} \simeq 0.45$, and that this evolution is fast, in the sense that a system should reside near $e_{\rm eq}$ following the accretion of only $\sim 1\%$ of its mass. Eccentric binaries are found to exhibit rich temporal accretion signals, showing distinct wave forms, pulse structures, and duty cycles in different eccentricity regimes. Here we address a few key applications of these results to binary stars and protostars, and massive black hole binaries. Details of the physical mechanism of gas-driven eccentricity evolution and further applications will be reported in \cite{Zrake2021}.

Our simulation setup is briefly summarized in \S\ref{sec:simulations}. In \S\ref{sec:eccentricity}, we report the computed eccentricity evolution as a function of $e$. We also provide an empirical fitting formula $\dot e(e)$ that can be used in modeling of binary populations. In \S\ref{sec:accretion-signatures} we present the accretion signatures of binaries at a range of eccentricities. In \S\ref{sec:gravitational-wave-signatures} we discuss gravitational wave signatures of eccentric binaries, and predict the eccentricity of massive black hole binaries entering the \emph{LISA} band. We summarize our results in \S\ref{sec:summary}.

\begin{figure}
\includegraphics{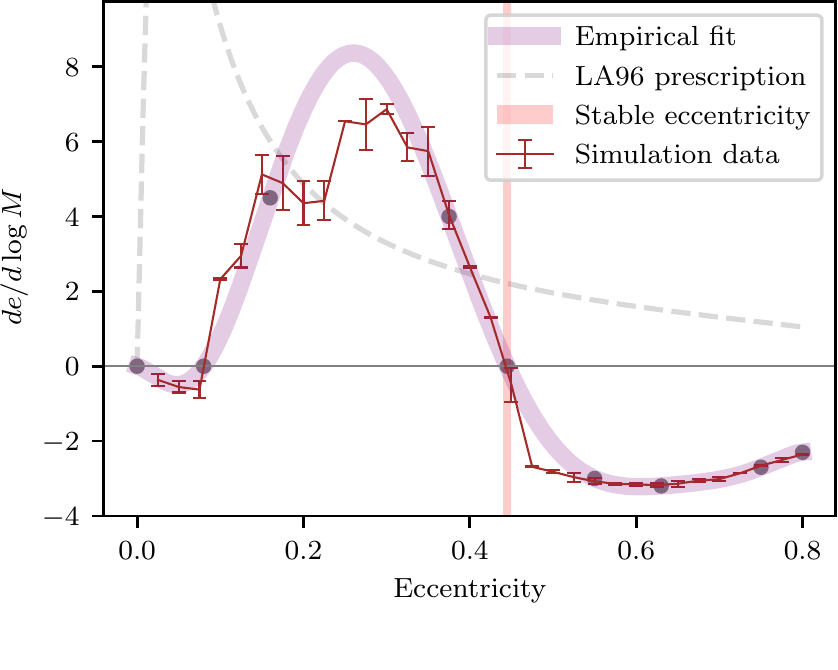}
\includegraphics{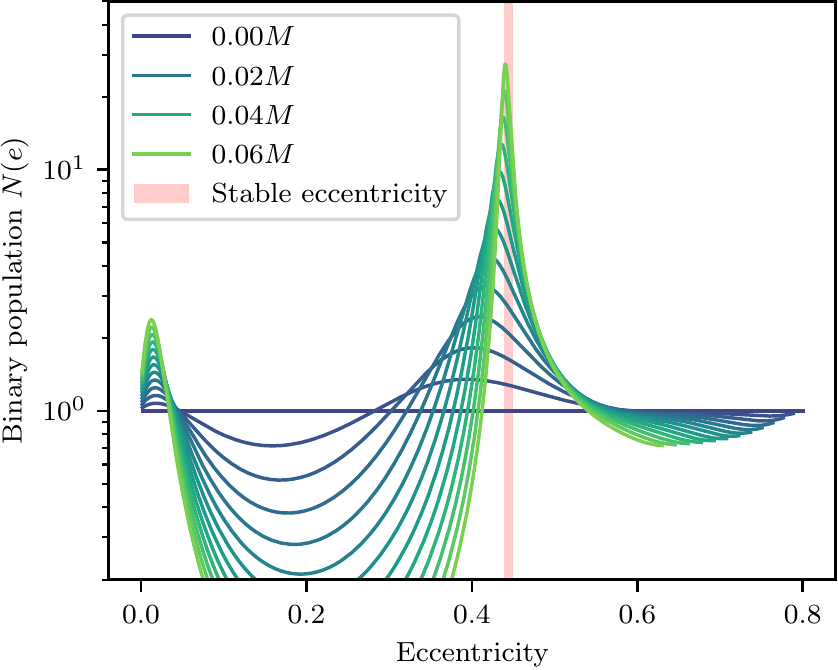}
\caption{\emph{\textbf{Top}} --- The rate of eccentricity evolution, per relative accreted mass, as a function of the binary eccentricity. Uncertainties are obtained by comparing two suites of 32 runs each, which are identical except for their numerical resolution. The purple band shows an empirical fit given by Eq. \ref{eqn:lagrange-polynomial}, and the grey circles show the control points from in Table \ref{tab:control-points}. The light dashed curve shows a commonly used prescription for $\dot e(e)$ from \citet{Lubow1996}. \emph{\textbf{Bottom}} --- Gas-driven evolution of a binary population $N(e)$, after accretion of given fractions (up to 6\%) of the binary mass $M$. The evolution is determined by using the empirical fit shown in the top panel.}
\label{fig:edot-vs-e}
\end{figure}

%
\section{Simulation setup} \label{sec:simulations}
We simulated the coupled evolution of an eccentric, equal-mass binary with a circumbinary gas disk. The disk is thin, two-dimensional, and locally isothermal with orbital Mach number $(h/r)^{-1} = 10$. Viscous stress is included with an $\alpha$-viscosity prescription with $\alpha=0.1$. Accretion onto the binary components is modeled by subtracting gas in a circular region of radius $r_{\rm sink} = 0.02a$, where $a$ is the binary semi-major axis. We utilize an initial condition in which the disk has a finite extent, and is thus free to expand outwards as it relaxes under the viscous stress. This setup captures binary evolution in the quasi-steady relaxed state of the disk, as well as the slow secular depletion of the surface density \citep{Munoz2020}. The simulation setup and hydrodynamic equations of motion are described fully in \cite{Tiede2020}.

We ran 64 simulations of eccentric binaries: low and high-resolution suites of 32 runs each, with eccentricities ranging from $0.025$ to $0.8$. The low-resolution runs had grid spacing $\Delta x \simeq 0.016a$, and the high-resolution runs had $\Delta x \simeq 0.012a$. Numerical convergence of the result was established by comparing the two suites, as shown in the error bars in Fig. \ref{fig:edot-vs-e}. Each run was evolved for three viscous relaxation times $t_{\rm visc}$, corresponding to roughly 2000 orbits. Stationarity was confirmed by inspecting the time series of the total mass accretion rate $\dot M$, the torque $\dot L$, and work $\dot E$ done by the disk on the binary. $\dot L$ and $\dot E$ include both the gravitational and accretion forces. The rate of change of the binary eccentricity is given by
\begin{equation}\label{eqn:edot}
  \frac{\dot e e}{1 - e^2} = \frac{\dot M}{M} + \frac{3 \dot \mu}{2\mu} - \frac{\dot E}{2E} - \frac{\dot L}{L} \, ,
\end{equation}
where $\mu \equiv M_1 M_2 / M$ is the reduced mass and $M_1$ and $M_2$ are the component masses. The binary mass $M \equiv M_1 + M_2$ is much larger than the mass accreted per disk relaxation time, consistent with the orbital eccentricity being fixed in each simulation. The gravitational and accretion forces which comprise the time derivatives on the right side of Eq. \ref{eqn:edot}, all scale linearly with $\dot M$. Thus, measurements of eccentricity evolution, which we report in terms of $\langle \dot e \rangle / \langle \dot M \rangle \equiv de/dM$, are valid in the limit $t_{\rm visc} \dot M \ll M$. Time averages $\langle \cdot \rangle$ were computed from $t = t_{\rm visc}$ to $3 t_{\rm visc}$.

%
\section{Eccentricity evolution} \label{sec:eccentricity}
Fig. \ref{fig:edot-vs-e} shows the rate of eccentricity change, per relative accreted mass ($de/d\log M$), as a function of the eccentricity. The four simulations at low eccentricity $e = 0.025$ through $e = 0.075$ exhibit $\dot e < 0$, and these binaries would thus be circularized by their interaction with the disk. Somewhere in the range $e = 0.075 - 0.1$, $\dot e$ increases abruptly and becomes positive. Eccentricity driving then increases in strength to a maximum $de/d\log M \simeq 7$ in the run with $e = 0.3$. The rate of eccentricity driving then smoothly declines, becoming negative between $e = 0.425$ and $e = 0.45$. The zero-crossing of $\dot e$ represents the stable eccentricity $e_{\rm eq}$, since binaries to the left of $e_{\rm eq}$ evolve to the right, and those to the right of $e_{\rm eq}$ evolve to the left.

The essential characteristics of the function $\dot e(e)$ are well captured by the empirical fit, shown in the light purple band in Fig. \ref{fig:edot-vs-e},
\begin{equation} \label{eqn:lagrange-polynomial}
\frac{de}{d \log M} \simeq
\sum_{j=0}^{8} \left(\frac{de}{d\log M} \right)_{\! \! j} \, \, \prod^8_{\shortstack{$\scriptstyle k=0 $\\$\scriptstyle j \ne k$}} \frac{e - e_k}{e_j - e_k} ,
\end{equation}
where the control points $e_j$ and $(de/d\log M)_j$ define a Lagrange interpolating polynomial and are given in Table \ref{tab:control-points}. An equivalent formula in terms of polynomial coefficients $a_j$, also given in Table \ref{tab:control-points}, is
\begin{equation} \label{eqn:direct-polynomial}
\frac{de}{d \log M} \simeq \sum_{j=0}^8 a_j e^j \, .
\end{equation}

To demonstrate the utility of the empirical fit, we use it to evolve the distribution $N(e)$ of a sample population of accreting binaries forward in time. The bottom panel of Fig. \ref{fig:edot-vs-e} shows this evolution, starting from a flat initial distribution $N(e) = {\rm constant}$ over the range $0 < e < 0.8$ and assuming that the accretion rate does not depend on $e$. After accreting a few percent of the binary mass the eccentricity distribution develops a pronounced peak around the equilibrium eccentricity $e_{\rm eq} \simeq 0.45$. A secondary peak of circular binaries also emerges, populated by the binaries starting with $e \lesssim 0.08$. Few binaries remain with intermediate eccentricities in the range $0.08 \lesssim e \lesssim 0.45$ and binaries initialized with $e \gtrsim 0.45$ are gradually depleted as well.

\subsection{Eccentricity of post-AGB stellar binaries}
\begin{figure}
\includegraphics{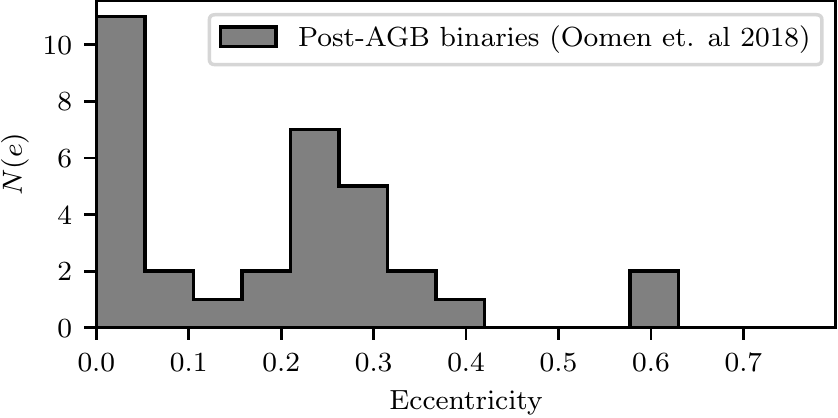}
\caption{Distribution of the eccentricity measurements of the 32 post-AGB stellar binaries reported in \citet{Oomen2018}. Each system has a period of $\lesssim 10$ years and should thus have been tidally circularized in the AGB phase.}
\label{fig:stellar-eccentricity}
\end{figure}

The computed eccentricity evolution $\dot e(e)$ may help explain the surprisingly high eccentricity of many post-AGB stellar binaries. These stars have previously ejected their hydrogen envelopes, and should have circularized via mass transfer in dense winds or Roche lobe overflow as they moved off the main sequence, provided their orbital periods are $\lesssim 10$ years \citep{Pols2003}. Post-AGB binaries on close orbits are nevertheless observed to have high eccentricities up to $e \approx 0.45$ \citep{Jorissen1998, Oomen2018}, as shown in Fig. \ref{fig:stellar-eccentricity}.

Tidal interaction with a circumbinary disk has been suggested to account for the eccentricity of the post-AGB binaries \citep{Waters1998}. Several studies, including \cite{Dermine2013}, \cite{Antoniadis2014}, and \cite{Oomen2018}, have modeled gas-driven eccentricity evolution using a prescription for $\dot e(e)$ suggested in \cite{Lubow1996}. This prescription, which we denote as $\dot e_{\rm LA}(e)$, is shown for comparison with our results in Fig. \ref{fig:edot-vs-e}. Note that since $\dot e_{\rm LA}(e) > 0$ for all $e$ including as $e \rightarrow 0$, the \cite{Lubow1996} prescription predicts spontaneous excitation of eccentricity from the well-circularized orbits expected in late-state AGB binaries, and is thus incompatible with the observed abundance of accreting and yet nearly circular binaries. In contrast, our results can account for the bi-modality in $N(e)$, as evidenced by the similarity between the simple predictions in Fig.\ref{fig:edot-vs-e} and the observed distribution in Fig.~\ref{fig:stellar-eccentricity},
but only if binaries are somehow perturbed onto slightly non-circular orbits by some other means. In other words, if the eccentricity distribution shown in Fig. \ref{fig:stellar-eccentricity} is indeed reflective of gas driven binary evolution, it leaves open the question of how the high-$e$ systems were nudged above the $e \simeq 0.08$ threshold.

\begin{figure*}
\includegraphics{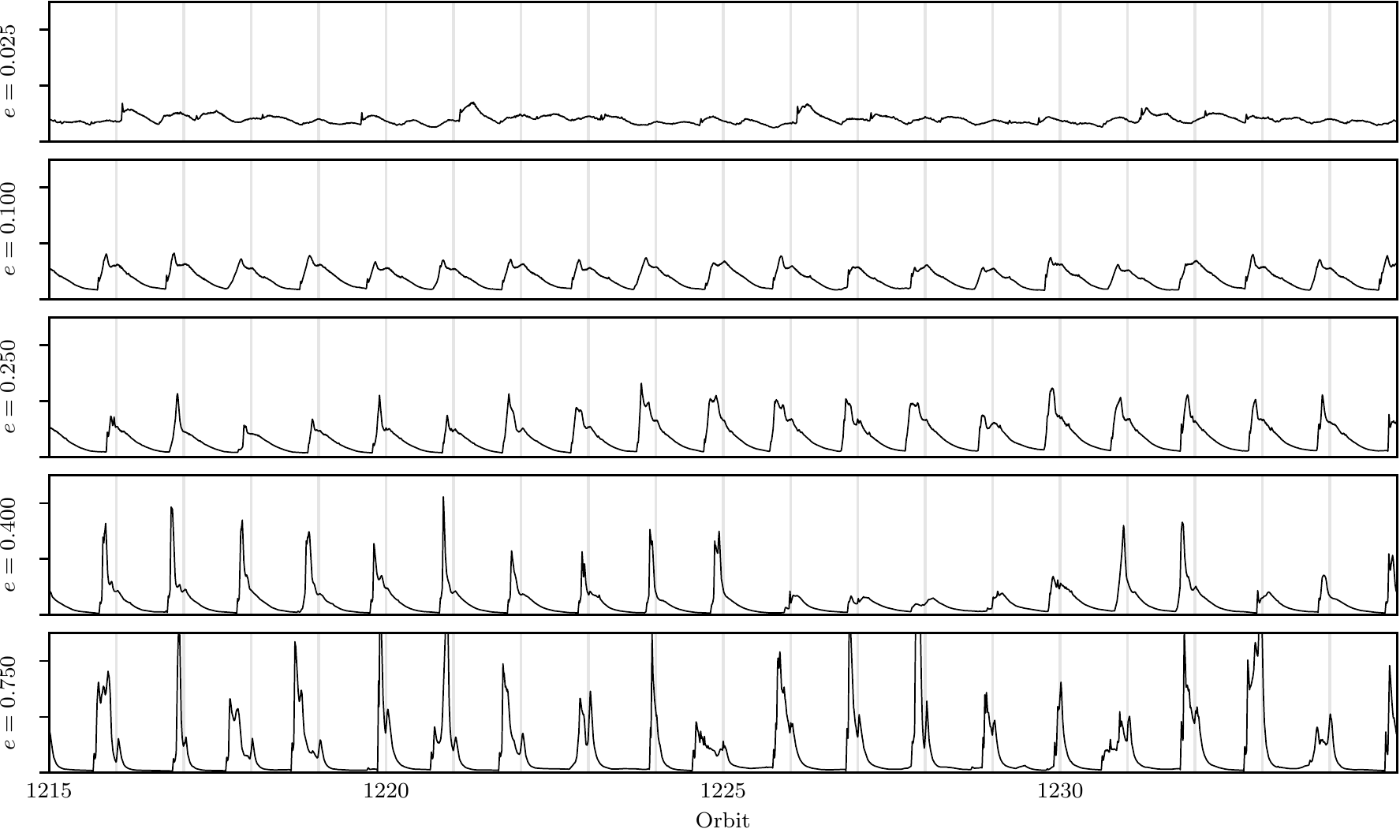}
\caption{Time series of the total accretion rate $\dot M$ at representative binary eccentricities: $e=0.025, \, 0.1, \, 0.25, \, 0.4, \, 0.75$ from top to bottom. The light vertical lines indicate periapse passages.}
\label{fig:accretion-time-series}
\end{figure*}

%
\section{Accretion signatures} \label{sec:accretion-signatures}
Fig. \ref{fig:accretion-time-series} shows the total accretion rate $\dot M$ as a function of time, over 20-orbit time windows, for binaries with increasing eccentricity. The $e=0.025$ case shown in the top-most panel exhibits features which are now well-established for circular equal-mass binaries \citep{MacFadyen2008, Farris2014, Munoz2019, Duffell2020}: $\dot M$ is modulated relatively smoothly in the orbital phase, with excursions of $\sim 25\%$ above and below the mean accretion rate. The enhancement in $\dot M$ every $\sim 5$ orbits is due to an $m=1$ density structure (or ``lump'') which is itself on an eccentric orbit around the binary at $r \sim 3a$, and transports a surplus of mass onto the binary at each periapse passage.

At $e=0.1$ the accretion rate develops a distinct saw-tooth pattern. Most orbits exhibit two local maxima in $\dot M$, the first preceding periapse by roughly 1/5 of an orbit, and the other occurring precisely at periapse. The first of the two peaks is usually of larger amplitude. Note that the saw-tooth pattern shows good regularity between orbits, and that the 5-orbit accretion enhancement is no longer present at $e=0.1$. This is due to suppression of the disk eccentricity, which is reported in detail in \cite{Zrake2021}, and is consistent with \cite{Miranda2017} who observed disappearance of the lump at $e=0.1$. At $e=0.25$ the saw-tooth pattern changes shape, now resembling a fast-rise exponential-decay type pulse, with the highest peak now always preceding periapse by a fraction of an orbit. The waveform develops modest irregularity, varying in amplitude somewhat from one orbit to the next. This trend continues to $e=0.4$; the accretion is now strongly concentrated in the pre-periapse spike, and the waveform develops increasing irregularity between orbits. At $e=0.75$ the waveform is even noisier and more irregular, exhibiting multiple accretion spikes during most orbits. The peak-to-trough ratio of $\dot M$ increases from $\sim 2$ for near-circular orbits to $\sim 100$ for $e \gtrsim 0.4$.

\subsection{Accretion in binary T-Tauri stars}
Our computed accretion signatures for binaries at a range of eccentricities may shed light on modulated and pulsed accretion observed in T-Tauri stellar binaries. For example, visual inspection of Fig. 9 from \cite{Jensen2007} suggests that modulating light curves of UZ Tau ($e\simeq 0.14-0.33$) might be a better fit to the saw-tooth pattern we see at $e \simeq 0.25$, than the sinusoidal wave form obtained by \cite{Artymowicz1996} from SPH simulations of an $e=0.1$ binary, which was the best simulation data those authors had available at that time. \cite{Martin2005} reported observations of UZ Tau in which the brightest peaks occured at orbital phase $0.88$, even though several observations were made closer to periastron, an effect which may reflect our computed signatures at $e = 0.25$ and $e = 0.4$ in which the peak accretion rate systematically precedes periastron by $10-20\%$ of an orbit.

The increasing irregularity of the accretion signal we see in our simulations at $e\gtrsim 0.4$ appears consistent with the incomplete duty cycle of pulsed emission from the high eccentricity system DQ Tau \citep[$e \simeq 0.57$;][]{Muzerolle2019}. \cite{Mathieu1997} reported significant non-detections of accretion pulses in as many as 35\% of pariastrons. Irregular light curves of DQ Tau were also reported in \cite{Kospal2018} and \cite{Tofflemire2017a}. \cite{Bary2014} observed an accretion flare around \emph{apastron}. Importantly, accretion pulses at the half-orbital period do begin to emerge in our simulations, infrequently at $e \simeq 0.55$ and becoming common toward $e=0.8$ (see orbits 1217, 1218, and 1224 in the bottom panel of Fig. \ref{fig:accretion-time-series}). \cite{Munoz2016} had speculated that the half-orbit accretion pulses, \emph{not} evident in their simulations of a binary with $e=0.5$, might have been numerically unresolved. However, observations of the high-$e$ system DQ Tau, together with simulations reported here, suggest that half-period spikes emerge only at $e \gtrsim 0.55$.

\begin{figure}
\includegraphics{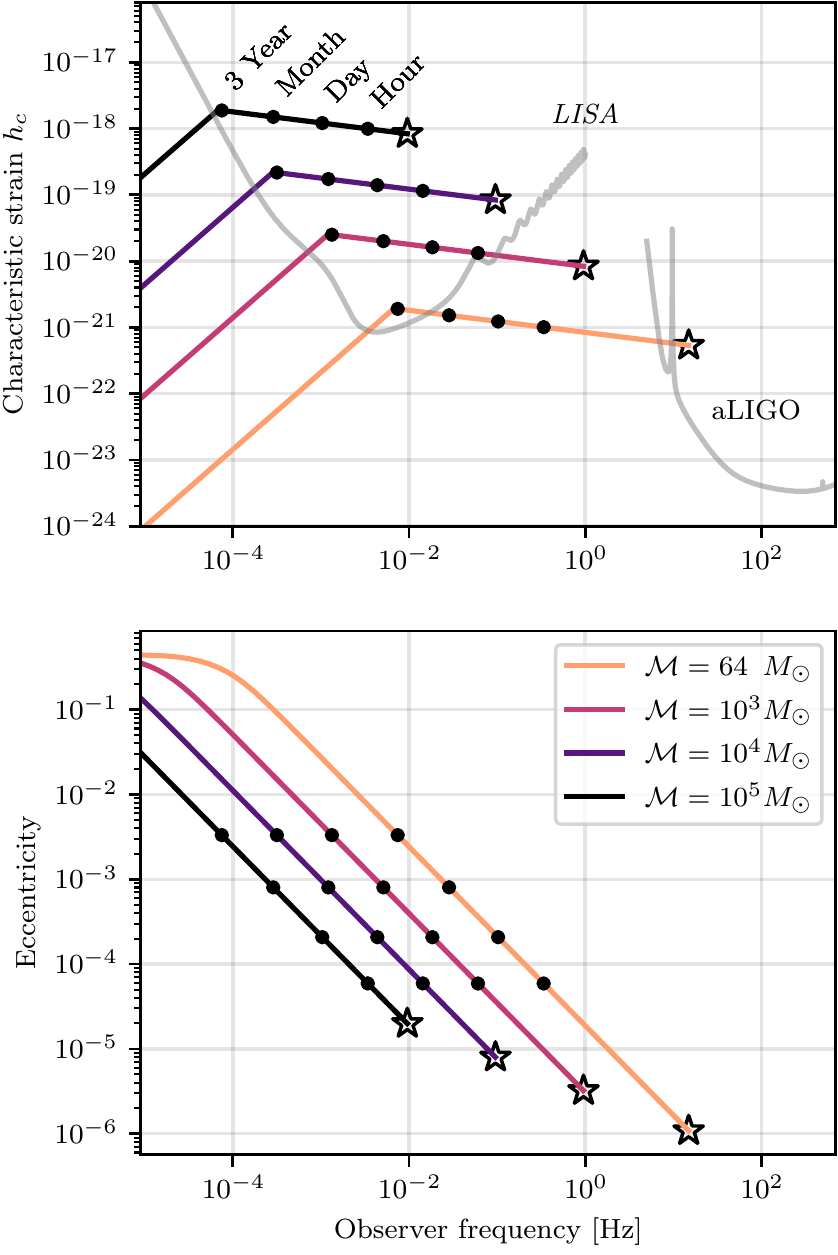}
\caption{\emph{\textbf{Top}} --- Characteristic strain of a sample of four different equal-mass black hole binaries, as they evolve through (observed) frequency $f$ in the gravitational-wave driven regime (note that $64 M_\odot$ is roughly the chirp mass associated with GW190521). The \emph{LISA} sensitivity curve $\sqrt{S_n f}$ is shown in grey \citep{Robson2019}. The sources are all located at redshift $z=1$. The time until merger is indicated in solid black circles, and star symbols mark the innermost stable circular orbit (ISCO). \emph{\textbf{Bottom}} --- Eccentricity versus observed frequency for the same sample black hole binaries. The binaries are initialized at $e=0.45$ near the equilibrium eccentricity obtained in the gas-dominated evolution phase, and circularize due to gravitational radiation as they evolve toward higher frequency.}
\label{fig:sensitivity-curve}
\end{figure}

\subsection{Electromagnetic signatures of MBHBs}
Our simulations indicate that MBHBs in the gas-driven regime (nominally at separations $\gtrsim \unit[0.01]{pc}$ for $10^{8-9} M_\odot$ BHs; e.g. \citealt{Haiman2009}) will have eccentricity $\sim e_{\rm eq}$. Optical variabilities of widely separated binary quasars in the gas-driven regime are thus expected to resemble the saw-tooth accretion patterns in the fourth panel of Fig. \ref{fig:accretion-time-series}, if their emission is dominated by accretion onto the black holes. More compact MBHBs evolving in the gravitational wave (GW) regime should, on the other hand, exhibit light curves characteristic of circular binaries. Dozens of binary quasar candidates have now been identified in large-scale optical time-domain surveys \citep[e.g.][]{Graham2015, Charisi2016, Chen2020}, most of which exhibit sinusoidally modulating light curves. If these systems are indeed MBHBs, they are either well into the GW-driven regime, or are exhibiting variability not directly connected to the accretion, such as Doppler modulations \citep{Dorazio+2015}. We also note that the algorithms used to search for periodicities among the large number of quasar light curves, based on Lomb-Scargle periodograms and their variants, are sensitive primarily to quasi-sinusoidal variability, and may miss the bursty or saw-tooth like periodicities. This should motivate the use of different search algorithms to mitigate a selection bias against eccentric binaries.

An intriguing exception is the periodic quasar candidate J0252 reported recently by \citealt{Liao2020}. This system exhibits significantly non-sinusoidal variability, which was interpreted in that study as evidence for a binary of unequal masses, by comparing its light curve to the simulated accretion signatures from \cite{Farris2014} of a binary with $M_2/M_1 = 0.11$. Although the light curves appear rather saw-tooth, as if they might also fit a mildly eccentric binary, we find the unequal mass scenario is indeed more likely, since J0252 is separated by $\lesssim \unit[10^{-3}]{pc}$ and almost certainly in the GW-driven regime. A systematic comparison between the computed light curves of eccentric versus unequal mass systems could help to differentiate otherwise ambiguous binary quasars.

The blazar OJ287, which exhibits a 12-year optical period, has been postulated to be an eccentric binary black hole. Our results here are not directly applicable to the ``standard model'' of this source, which involves an unequal-mass binary on an orbit tilted with respect to the circumbinary disk, and fits its periodic light-curve, with recurring double peaks (see, e.g., \citealt{Laine2020} and \citealt{Dey2018} and references therein). While our results suggest that an eccentric binary naturally arises even in the co-planar case, the observed regularity of the OJ287 pulses seems incompatible with our computed accretion signals for eccentric binaries.

%
\section{Gravitational wave signatures} \label{sec:gravitational-wave-signatures}
\begin{figure}
\includegraphics{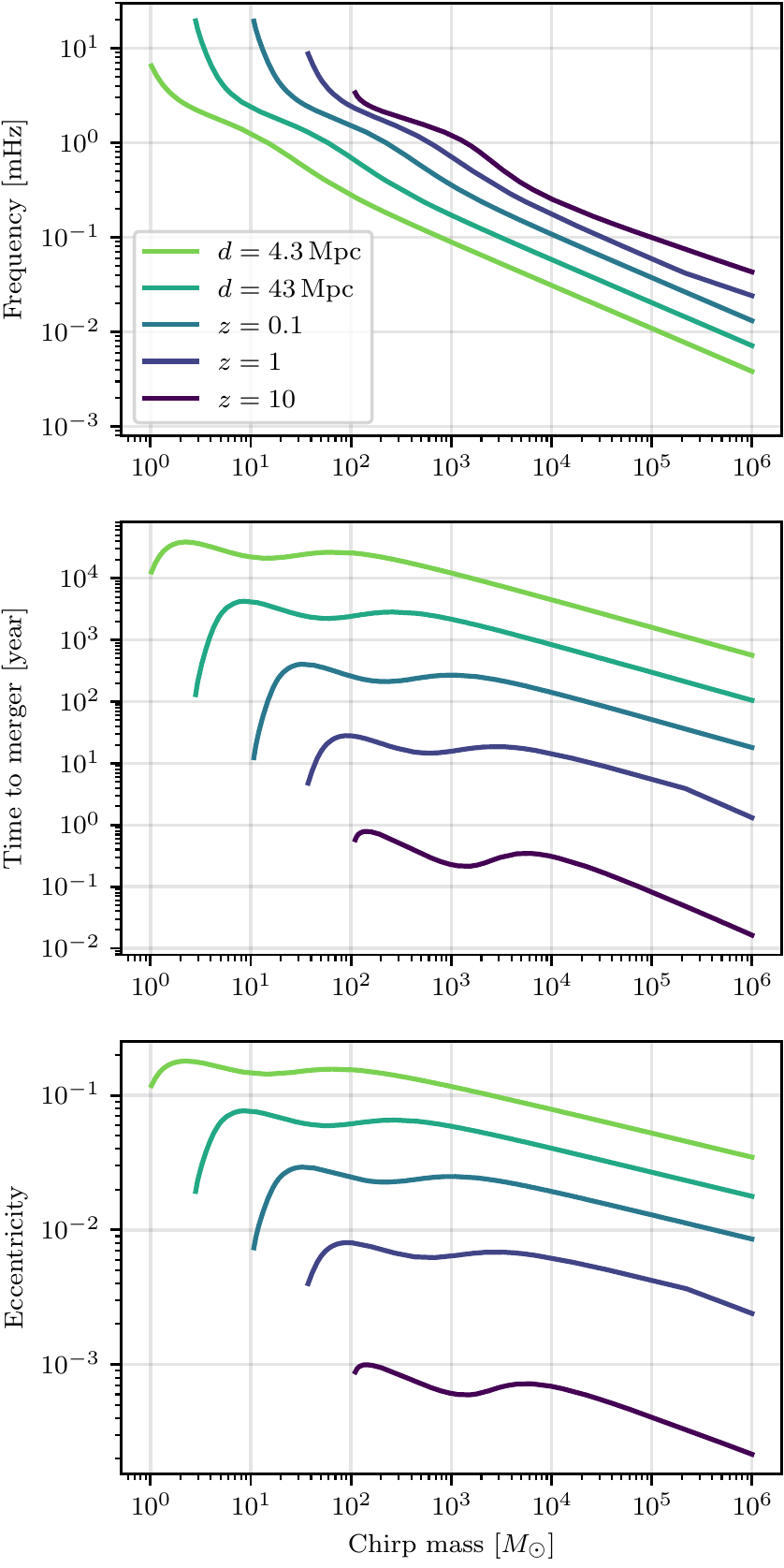}
\caption{Binary parameters upon entering the \emph{LISA} band, as a function of the chirp mass $\mathcal{M}$, for sample redshifts from $z=10$ (\emph{dark}) down to $z=10^{-3}$ (\emph{light}). The frequency (\emph{top}), merger time (\emph{middle}), and eccentricity (\emph{bottom}) are at the time when the binary characteristic strain $h_c$ first intersects the \emph{LISA} sensitivity curve as shown in Fig. \ref{fig:sensitivity-curve}.}
\label{fig:entering-lisa-band}
\end{figure}

\subsection{LISA}
Massive black holes (MBHBs) with $M\approx 10^{2-7} M_\odot$ produce gravitational waves (GWs) that could be detected by the planned space interferometer \emph{LISA} in the days to years leading up to their coalescence \citep[e.g.][]{LISA}. \emph{LISA} is expected to be sensitive to source eccentricities as small $e \sim 10^{-3}$, and it has been argued \citep{Armitage2005, Cuadra2009} that residual eccentricity from a gas-driven episode might be imprinted on MBHBs in the \emph{LISA} band. Here we update and expand on those calculations, taking into account our computed equilibrium eccentricity $e_{\rm eq} \simeq 0.45$, updated \emph{LISA} sensitivity curves from \cite{Robson2019}, and a wider range of source masses and redshifts.

\citet{MilosPhinney2005} proposed that beyond the so-called decoupling radius, at which the viscous timescale in the nearby disk exceeds the GW inspiral time, the binary runs away without the disk being able to follow. However, \citet{Farris+2015} and \citet{Tang+2018} found that angular momentum transfer by shocks occurs on the orbital timescale, enabling the inner disk to follow the rapidly inspiraling binary. We therefore include the disk torques throughout the inspiral, in contrast with prior works which assumed that disk torques are abruptly suppressed at the nominal decoupling radius \citep[e.g.][]{Armitage2005, Roedig2011}.

The notion of a decoupling radius is only valid if binaries are hardened ($\dot a < 0$) by gas driving, but this need not always be the case: circular equal-mass binaries surrounded by $h/r = 0.1$ disks absorb angular momentum from the disk and are instead widened \citep{Tang2017, Miranda2017, Munoz2019, Tiede2020, Duffell2020}. However, in \cite{Zrake2021} we show that binaries at the stable eccentricity experience gas-driven \emph{hardening}, at a rate $d\log a / d\log M \sim -1$.

We therefore initialize widely-separated binaries in a gas-driven phase at the stable eccentricity $e_{\rm eq}$, and evolve them as they inspiral due to the combined effects of the gas disk and GW emission. Given Eq. \ref{eqn:lagrange-polynomial} for $de / d\log M$, and the assumption that binaries accrete at the Eddington rate, we obtain equations for $\dot a_{\rm gas}$ and $\dot e_{\rm gas}$. We then add the post-Newtonian terms for $\dot a_{\rm GW}$ and $\dot e_{\rm GW}$ from \cite{Peters1964} to account for gravitational wave driving, and solve the resulting differential equation for the combined influence of gas and GWs numerically.

Fig. \ref{fig:sensitivity-curve} shows the result of this calculation, for sample binaries initiated in the gas-dominated phase with $e = 0.45$. The top panel shows the trajectory in characteristic strain $h_c$ versus observed frequency $f$, for four representative binaries of different chirp masses, and assuming a 4-year observation time (i.e. \emph{LISA} mission lifetime). The bottom panel shows the trajectory of the same sample binaries as they evolve through eccentricity and observed frequency. Each system will be detectable by \emph{LISA} at least 3 years before it merges, and at that time its eccentricity is $0.003$. Thus we predict that 3 years prior to merger, \emph{LISA} binaries delivered to the GW regime by gas accretion will have a measurable eccentricity $\gtrsim 10^{-3}$.

Binaries just entering the \emph{LISA} band will have still higher eccentricity. Fig. 5 shows, as a function of the binary chirp mass and redshift, the orbital eccentricity at the time the source is first detectable (i.e. the characteristic strain is above \emph{LISA}'s noise). Note that very nearby sources within $\sim \unit[40]{Mpc}$ will have $e \sim 0.02 - 0.1$, depending on chirp mass. Sources at $z = 1$ with chirp masses $\mathcal{M} = 10^2 - 10^4 M_\odot$ are predicted to have $e = 0.005 - 0.01$. These values are larger by a factor of $\sim$few than those obtained in \cite{Armitage2005}, but compatible with the estimates of \cite{Cuadra2009}.

\subsection{PTAs}
Supermassive binary black hole ($10^{7-10} M_\odot$) systems generate a stochastic background of low-frequency GWs, which may be consistent with the signal recently measured in NANOGrav's 12.5yr data \citep{NANOGravGWB} and is expected to be confidently detected by further pulsar timing array (PTA) campaigns over the next several years \citep{Mingarelli2019}. Gas disks are known to significantly impact the unresolved stochastic background \citep{KocsisSesana2011}. For example, eccentricity tilts the single-source spectrum \citep{Sesana2015} by shifting GW power to higher harmonics. Eccentric binaries also harden faster than circular ones, reducing the number of background sources around the decoupling frequency \citep{Sesana2015}. PTAs are also expected to resolve individual sources, and these are likely to be near the fiducial decoupling stage \citep{Kelley2017}. Our results therefore have the potentially significant implication that individual PTA binaries will be highly eccentric, with $e$ at, or not far below, its equilibrium value.

\subsection{LIGO/Virgo}
Fig. \ref{fig:sensitivity-curve} shows that if binaries like GW190521, with chirp mass $\sim 64 M_\odot$ and at redshift $z\sim 1$, accrete from circumbinary disks, they will cross through the \emph{LISA} band with detectable eccentricity $e \sim 5\times 10^{-3}$. Such events merge in the aLIGO band, however upon entering it their eccentricity is further suppressed by GW circularization to $e \sim 10^{-6}$. Therefore, the large eccentricities $e \gtrsim 0.1$ and $e\sim 0.7$ suggested in \cite{Romero-Shaw+2020} and \cite{Gayathri+2020} respectively cannot be attributed to a circumbinary disk as discussed in this {\it Letter}.

%
\section{Summary} \label{sec:summary}
We have examined the eccentricity evolution of equal-mass binary systems accreting from a circumbinary gas disk, mapping out the rate of eccentricity change $\dot e$ as a function of $e$ in the range $e = 0.025 - 0.8$. Based on a suite of simulations where the disk has aspect ratio $h/r = 0.1$ and viscosity parameter $\alpha=0.1$, binaries with $e \lesssim 0.08$ are circularized, and those with $e \gtrsim 0.08$ are driven toward an equilibrium eccentricity $e_{\rm eq} \simeq 0.45$. The evolution is fast, in the sense that the binary population develops an order-unity enhancement around $e_{\rm eq}$ after systems have accreted a mere $\sim 1\%$ of their mass.

We provided an empirical fit $\dot e(e)$ to our simulation results, for use in population synthesis modeling of accreting black hole and stellar binaries. It predicts an eccentricity distribution $N(e)$ that is bi-modal, and highly suggestive of the population of accreting post-AGB stellar binaries. However, insofar as these systems ought to have been circularized by mass transfer at an earlier stage, they should have remained circular given the modest threshold $e \simeq 0.08$ for eccentricity growth. It suggests either an unknown mechanism to halt circularization in the AGB phase \cite[e.g.][]{Marinovic2008}, or that the eccentricity of post-AGB stars is not ultimately controlled by accretion from the circumbinary disk \citep{Rafikov2016}.

The temporal signatures of eccentric binary accretion are strikingly diverse. The 5-orbit periodicity occurring in circular binaries is suppressed for $e \gtrsim 0.1$. In the range $e \simeq 0.1 - 0.4$, the periodicity develops a distinct saw-tooth shape, and accretion spikes typically precede periapse passage in orbital phase by $10-20\%$. At $e \gtrsim 0.4$ the accretion becomes increasingly irregular, and the pulses vary dramatically in amplitude from one orbit to the next. Around $e \gtrsim 0.55$, the accretion pulses begin to spread out in orbital phase, sometimes even occurring around apoapse. We discussed these accretion signals in the context of the spectroscopic proto-stellar binaries UZ Tau and DQ Tau, and found several consistencies with our simulations, including with the orbital phase of the pulses, and their increasing irregularity at high $e$.

We discussed prospects for the application of eccentric accretion signals for interpreting the light curves of binary quasar candidates. Assuming that the optical variability of these systems reflects BH accretion rate modulations, we predict that the variability is saw-tooth in the gas-driven regime ($e \simeq e_{\rm eq}$), and exhibits the well-known 5-orbit periodicity only in the GW-driven regime after the orbit has largely circularized.

Finally, we showed that MBHBs that are delivered to the GW regime through binary accretion are likely to enter the \emph{LISA} band with a measurable eccentricity $10^{-2} - 10^{-3}$. This applies to \emph{LISA} sources of chirp mass as low as $\sim 10^2 M_\odot$ and as high as $10^4 M_\odot$ out to a redshift $z \sim 10$, or within $z = 1$ and with chirp mass as large as $10^6 M_\odot$. Comparable-mass PTA sources detected around the gas decoupling stage are predicted to be eccentric with $e \simeq 0.4 - 0.5$.

%
\newpage
\appendix
Table \ref{tab:control-points} contains the coefficients used in either of the equivalent empirical fit formulas (Eq. \ref{eqn:lagrange-polynomial} or \ref{eqn:direct-polynomial}) for $de/d\log M$. A Python function to evaluate Eq. \ref{eqn:lagrange-polynomial} is returned by \texttt{scipy.interpolate.BarycentricInterpolator(ej, yj)}, where \texttt{ej} and \texttt{yj} are the two middle columns of Table \ref{tab:control-points}.

\newcommand\Tstrut{\rule{0pt}{2.75ex}}         
\newcommand\Bstrut{\rule[-1.25ex]{0pt}{0pt}}   

\begin{table}[h!]
\centering
\begin{tabular}{l|llc}
\hline
\hline
$j$ & $e_j$ & $(de/d\log M)_j$ & $a_j$ \Tstrut\Bstrut \\
\hline
0 & 0.000 & $+0.0$ & $+0.0000e0$ \\
1 & 0.080 & $+0.0$ & $-5.5122e0$ \\
2 & 0.160 & $+4.5$ & $-5.5540e2$ \\
3 & 0.375 & $+4.0$ & $+1.2667e4$ \\
4 & 0.445 & $+0.0$ & $-7.5392e4$ \\
5 & 0.550 & $-3.0$ & $+2.0419e5$ \\
6 & 0.630 & $-3.2$ & $-2.8660e5$ \\
7 & 0.750 & $-2.7$ & $+2.0393e5$ \\
8 & 0.800 & $-2.3$ & $-5.8380e4$ \\
\hline
\end{tabular}
\caption{Table of control points used in the empirical fitting relation between the binary eccentricity $e$ rate of eccentricity driving $de/d\log M$ shown in Fig. \ref{fig:edot-vs-e}. The first two columns define an 8th degree Lagrange interpolating polynomial, shown in Eq. \ref{eqn:lagrange-polynomial}, and the final column contains the equivalent coefficients written in the direct polynomial basis, Eq. \ref{eqn:direct-polynomial}.}
\label{tab:control-points}
\end{table}

%
\acknowledgments
J. Zrake acknowledges S. Brittain for valuable input. The authors thank Daniel D'Orazio and John-Ryan Westernacher-Schneider for comments on the manuscript.
Resources supporting this work were provided by the NASA High-End Computing (HEC) Program through the NASA Advanced Supercomputing (NAS) Division at Ames Research Center.
We acknowledge support from NASA grant NNX15AB19G, and NSF grants AST-2006176 and  AST-1715661.

\bibliography{refs}

\begin{thebibliography}{}
\expandafter\ifx\csname natexlab\endcsname\relax\def\natexlab#1{#1}\fi
\providecommand{\url}[1]{\href{#1}{#1}}

\bibitem[{{Alves} {et~al.}(2019){Alves}, {Caselli}, {Girart}, {Segura-Cox},
  {Franco}, {Schmiedeke}, \& {Zhao}}]{Alves2019}
{Alves}, F.~O., {Caselli}, P., {Girart}, J.~M., {et~al.} 2019, Science, 366, 90

\bibitem[{{Amaro-Seoane} {et~al.}(2017){Amaro-Seoane}, {Audley}, {Babak},
  {Baker}, {Barausse}, {Bender}, {Berti}, {Binetruy}, {Born}, {Bortoluzzi},
  {Camp}, {Caprini}, {Cardoso}, {Colpi}, {Conklin}, {Cornish}, {Cutler},
  {Danzmann}, {Dolesi}, {Ferraioli}, {Ferroni}, {Fitzsimons}, {Gair}, {Gesa
  Bote}, {Giardini}, {Gibert}, {Grimani}, {Halloin}, {Heinzel}, {Hertog},
  {Hewitson}, {Holley-Bockelmann}, {Hollington}, {Hueller}, {Inchauspe},
  {Jetzer}, {Karnesis}, {Killow}, {Klein}, {Klipstein}, {Korsakova}, {Larson},
  {Livas}, {Lloro}, {Man}, {Mance}, {Martino}, {Mateos}, {McKenzie},
  {McWilliams}, {Miller}, {Mueller}, {Nardini}, {Nelemans}, {Nofrarias},
  {Petiteau}, {Pivato}, {Plagnol}, {Porter}, {Reiche}, {Robertson},
  {Robertson}, {Rossi}, {Russano}, {Schutz}, {Sesana}, {Shoemaker}, {Slutsky},
  {Sopuerta}, {Sumner}, {Tamanini}, {Thorpe}, {Troebs}, {Vallisneri},
  {Vecchio}, {Vetrugno}, {Vitale}, {Volonteri}, {Wanner}, {Ward}, {Wass},
  {Weber}, {Ziemer}, \& {Zweifel}}]{LISA}
{Amaro-Seoane}, P., {Audley}, H., {Babak}, S., {et~al.} 2017, Proposal
  submitted to ESA on January 13th, 2007; e-print arXiv:1702.00786

\bibitem[{{Antoniadis}(2014)}]{Antoniadis2014}
{Antoniadis}, J. 2014, \apjl, 797, L24

\bibitem[{{Armitage} \& {Natarajan}(2005)}]{Armitage2005}
{Armitage}, P.~J., \& {Natarajan}, P. 2005, \apj, 634, 921

\bibitem[{{Artymowicz}(1992)}]{Artymowicz1992}
{Artymowicz}, P. 1992, \pasp, 104, 769

\bibitem[{{Artymowicz} {et~al.}(1991){Artymowicz}, {Clarke}, {Lubow}, \&
  {Pringle}}]{Artymowicz1991}
{Artymowicz}, P., {Clarke}, C.~J., {Lubow}, S.~H., \& {Pringle}, J.~E. 1991,
  \apjl, 370, L35

\bibitem[{{Artymowicz} \& {Lubow}(1996)}]{Artymowicz1996}
{Artymowicz}, P., \& {Lubow}, S.~H. 1996, \apjl, 467, L77

\bibitem[{{Arzoumanian} {et~al.}(2020){Arzoumanian}, {Baker}, {Blumer},
  {Becsy}, {Brazier}, {Brook}, {Burke-Spolaor}, {Chatterjee}, {Chen}, {Cordes},
  {Cornish}, {Crawford}, {Cromartie}, {DeCesar}, {Demorest}, {Dolch}, {Ellis},
  {Ferrara}, {Fiore}, {Fonseca}, {Garver-Daniels}, {Gentile}, {Good},
  {Hazboun}, {Holgado}, {Islo}, {Jennings}, {Jones}, {Kaiser}, {Kaplan},
  {Kelley}, {Shapiro Key}, {Laal}, {Lam}, {Lazio}, {Lorimer}, {Luo}, {Lynch},
  {Madison}, {McLaughlin}, {Mingarelli}, {Ng}, {Nice}, {Pennucci}, {Pol},
  {Ransom}, {Ray}, {Shapiro-Albert}, {Siemens}, {Simon}, {Spiewak}, {Stairs},
  {Stinebring}, {Stovall}, {Sun}, {Swiggum}, {Taylor}, {Turner}, {Vallisneri},
  {Vigeland}, \& {Witt}}]{NANOGravGWB}
{Arzoumanian}, Z., {Baker}, P.~T., {Blumer}, H., {et~al.} 2020, \apjl,
  submitted; e-print arXiv:2009.04496

\bibitem[{{Baruteau} {et~al.}(2011){Baruteau}, {Cuadra}, \&
  {Lin}}]{Baruteau+2011}
{Baruteau}, C., {Cuadra}, J., \& {Lin}, D.~N.~C. 2011, \apj, 726, 28

\bibitem[{Bary \& Petersen(2014)}]{Bary2014}
Bary, J.~S., \& Petersen, M.~S. 2014, The Astrophysical Journal, 792, 64

\bibitem[{{Begelman} {et~al.}(1980){Begelman}, {Blandford}, \&
  {Rees}}]{Begelman1980}
{Begelman}, M.~C., {Blandford}, R.~D., \& {Rees}, M.~J. 1980, \nat, 287, 307

\bibitem[{{Bona{\v{c}}i{\'c} Marinovi{\'c}} {et~al.}(2008){Bona{\v{c}}i{\'c}
  Marinovi{\'c}}, {Glebbeek}, \& {Pols}}]{Marinovic2008}
{Bona{\v{c}}i{\'c} Marinovi{\'c}}, A.~A., {Glebbeek}, E., \& {Pols}, O.~R.
  2008, \aap, 480, 797

\bibitem[{{Charisi} {et~al.}(2016){Charisi}, {Bartos}, {Haiman},
  {Price-Whelan}, {Graham}, {Bellm}, {Laher}, \& {M{\'a}rka}}]{Charisi2016}
{Charisi}, M., {Bartos}, I., {Haiman}, Z., {et~al.} 2016, \mnras, 463, 2145

\bibitem[{{Chen} {et~al.}(2020){Chen}, {Liu}, {Liao}, {Holgado}, {Guo},
  {Gruendl}, {Morganson}, {Shen}, {Zhang}, {Abbott}, {Aguena}, {Allam},
  {Avila}, {Bertin}, {Bhargava}, {Brooks}, {Burke}, {Rosell}, {Carollo},
  {Kind}, {Carretero}, {Costanzi}, {da Costa}, {Davis}, {De Vicente}, {Desai},
  {Diehl}, {Doel}, {Everett}, {Flaugher}, {Friedel}, {Frieman},
  {Garc{\'\i}a-Bellido}, {Gaztanaga}, {Glazebrook}, {Gruen}, {Gutierrez},
  {Hinton}, {Hollowood}, {James}, {Kim}, {Kuehn}, {Kuropatkin}, {Lewis},
  {Lidman}, {Lima}, {Maia}, {March}, {Marshall}, {Menanteau}, {Miquel},
  {Palmese}, {Paz-Chinch{\'o}n}, {Plazas}, {Sanchez}, {Schubnell}, {Serrano},
  {Sevilla-Noarbe}, {Smith}, {Suchyta}, {Swanson}, {Tarle}, {Tucker}, {Varga},
  \& {Walker}}]{Chen2020}
{Chen}, Y.-C., {Liu}, X., {Liao}, W.-T., {et~al.} 2020, \mnras,
  arXiv:2008.12329

\bibitem[{{Cuadra} {et~al.}(2009){Cuadra}, {Armitage}, {Alexander}, \&
  {Begelman}}]{Cuadra2009}
{Cuadra}, J., {Armitage}, P.~J., {Alexander}, R.~D., \& {Begelman}, M.~C. 2009,
  \mnras, 393, 1423

\bibitem[{{Dermine} {et~al.}(2013){Dermine}, {Izzard}, {Jorissen}, \& {Van
  Winckel}}]{Dermine2013}
{Dermine}, T., {Izzard}, R.~G., {Jorissen}, A., \& {Van Winckel}, H. 2013,
  \aap, 551, A50

\bibitem[{{Dey} {et~al.}(2018){Dey}, {Valtonen}, {Gopakumar}, {Zola}, {Hudec},
  {Pihajoki}, {Ciprini}, {Matsumoto}, {Sadakane}, {Kidger}, {Nilsson},
  {Mikkola}, {Sillanp{\"a}{\"a}}, {Takalo}, {Lehto}, {Berdyugin}, {Piirola},
  {Jermak}, {Baliyan}, {Pursimo}, {Caton}, {Alicavus}, {Baransky}, {Blay},
  {Boumis}, {Boyd}, {Campas Torrent}, {Campos}, {Carrillo G{\'o}mez},
  {Chandra}, {Chavushyan}, {Dalessio}, {Debski}, {Drozdz}, {Er}, {Erdem},
  {Escartin P{\'e}rez}, {Fallah Ramazani}, {Filippenko}, {Gafton}, {Ganesh},
  {Garcia}, {Gazeas}, {Godunova}, {G{\'o}mez Pinilla}, {Gopinathan}, {Haislip},
  {Harmanen}, {Hurst}, {Jan{\'\i}k}, {Jelinek}, {Joshi}, {Kagitani},
  {Karjalainen}, {Kaur}, {Keel}, {Kouprianov}, {Kundera}, {Kurowski},
  {Kvammen}, {LaCluyze}, {Lee}, {Liakos}, {Lindfors}, {Lozano de Haro},
  {Mugrauer}, {Naves Nogues}, {Neely}, {Nelson}, {Ogloza}, {Okano},
  {Pajdosz-{\'S}mierciak}, {Pand ey}, {Perri}, {Poyner}, {Provencal}, {Raj},
  {Reichart}, {Reinthal}, {Reynolds}, {Saario}, {Sadegi}, {Sakanoi}, {Salto
  Gonz{\'a}lez}, {Sameer}, {Schweyer}, {Simon}, {Siwak}, {Sold{\'a}n Alfaro},
  {Sonbas}, {Steele}, {Stocke}, {Strobl}, {Tomov}, {Tremosa Espasa}, {Valdes},
  {Valero P{\'e}rez}, {Verrecchia}, {Vasylenko}, {Webb}, {Yoneda}, {Zejmo},
  {Zheng}, \& {Zielinski}}]{Dey2018}
{Dey}, L., {Valtonen}, M.~J., {Gopakumar}, A., {et~al.} 2018, \apj, 866, 11

\bibitem[{{D'Orazio} {et~al.}(2015){D'Orazio}, {Haiman}, \&
  {Schiminovich}}]{Dorazio+2015}
{D'Orazio}, D.~J., {Haiman}, Z., \& {Schiminovich}, D. 2015, \nat, 525, 351

\bibitem[{{Duffell} {et~al.}(2020){Duffell}, {D'Orazio}, {Derdzinski},
  {Haiman}, {MacFadyen}, {Rosen}, \& {Zrake}}]{Duffell2020}
{Duffell}, P.~C., {D'Orazio}, D., {Derdzinski}, A., {et~al.} 2020, \apj, 901,
  25

\bibitem[{{Farris} {et~al.}(2014){Farris}, {Duffell}, {MacFadyen}, \&
  {Haiman}}]{Farris2014}
{Farris}, B.~D., {Duffell}, P., {MacFadyen}, A.~I., \& {Haiman}, Z. 2014, \apj,
  783, 134

\bibitem[{{Farris} {et~al.}(2015){Farris}, {Duffell}, {MacFadyen}, \&
  {Haiman}}]{Farris+2015}
---. 2015, \mnras, 447, L80

\bibitem[{{Gayathri} {et~al.}(2020){Gayathri}, {Healy}, {Lange}, {O'Brien},
  {Szczepanczyk}, {Bartos}, {Campanelli}, {Klimenko}, {Lousto}, \&
  {O'Shaughnessy}}]{Gayathri+2020}
{Gayathri}, V., {Healy}, J., {Lange}, J., {et~al.} 2020, arXiv e-prints,
  arXiv:2009.05461

\bibitem[{{Goldreich} \& {Tremaine}(1980)}]{Goldreich1980}
{Goldreich}, P., \& {Tremaine}, S. 1980, \apj, 241, 425

\bibitem[{{Graham} {et~al.}(2015){Graham}, {Djorgovski}, {Stern}, {Drake},
  {Mahabal}, {Donalek}, {Glikman}, {Larson}, \& {Christensen}}]{Graham2015}
{Graham}, M.~J., {Djorgovski}, S.~G., {Stern}, D., {et~al.} 2015, \mnras, 453,
  1562

\bibitem[{{Haiman} {et~al.}(2009){Haiman}, {Kocsis}, \& {Menou}}]{Haiman2009}
{Haiman}, Z., {Kocsis}, B., \& {Menou}, K. 2009, \apj, 700, 1952

\bibitem[{{Jensen} {et~al.}(2007){Jensen}, {Dhital}, {Stassun}, {Patience},
  {Herbst}, {Walter}, {Simon}, \& {Basri}}]{Jensen2007}
{Jensen}, E. L.~N., {Dhital}, S., {Stassun}, K.~G., {et~al.} 2007, \aj, 134,
  241

\bibitem[{{Jorissen} {et~al.}(1998){Jorissen}, {Van Eck}, {Mayor}, \&
  {Udry}}]{Jorissen1998}
{Jorissen}, A., {Van Eck}, S., {Mayor}, M., \& {Udry}, S. 1998, \aap, 332, 877

\bibitem[{{Kelley} {et~al.}(2017){Kelley}, {Blecha}, \&
  {Hernquist}}]{Kelley2017}
{Kelley}, L.~Z., {Blecha}, L., \& {Hernquist}, L. 2017, \mnras, 464, 3131

\bibitem[{{Kocsis} \& {Sesana}(2011)}]{KocsisSesana2011}
{Kocsis}, B., \& {Sesana}, A. 2011, \mnras, 411, 1467

\bibitem[{K{\'{o}}sp{\'{a}}l {et~al.}(2018)K{\'{o}}sp{\'{a}}l,
  {\'{A}}brah{\'{a}}m, Zsidi, Vida, Szab{\'{o}}, Mo{\'{o}}r, \&
  P{\'{a}}l}]{Kospal2018}
K{\'{o}}sp{\'{a}}l, {\'{A}}., {\'{A}}brah{\'{a}}m, P., Zsidi, G., {et~al.}
  2018, The Astrophysical Journal, 862, 44

\bibitem[{{Laine} {et~al.}(2020){Laine}, {Dey}, {Valtonen}, {Gopakumar},
  {Zola}, {Komossa}, {Kidger}, {Pihajoki}, {G{\'o}mez}, {Caton}, {Ciprini},
  {Drozdz}, {Gazeas}, {Godunova}, {Haque}, {Hildebrand t}, {Hudec}, {Jermak},
  {Kong}, {Lehto}, {Liakos}, {Matsumoto}, {Mugrauer}, {Pursimo}, {Reichart},
  {Simon}, {Siwak}, \& {Sonbas}}]{Laine2020}
{Laine}, S., {Dey}, L., {Valtonen}, M., {et~al.} 2020, \apjl, 894, L1

\bibitem[{{Liao} {et~al.}(2020){Liao}, {Chen}, {Liu}, {Holgado}, {Guo},
  {Gruendl}, {Morganson}, {Shen}, {Davis}, {Kessler}, {Martini}, {McMahon},
  {Allam}, {Annis}, {Avila}, {Banerji}, {Bechtol}, {Bertin}, {Brooks},
  {Buckley-Geer}, {Carnero Rosell}, {Carrasco Kind}, {Carretero}, {Castander},
  {Cunha}, {D'Andrea}, {da Costa}, {Davis}, {De Vicente}, {Desai}, {Diehl},
  {Doel}, {Eifler}, {Evrard}, {Flaugher}, {Fosalba}, {Frieman},
  {Garcia-Bellido}, {Gaztanaga}, {Glazebrook}, {Gruen}, {Gschwend},
  {Gutierrez}, {Hartley}, {Hollowood}, {Honscheid}, {Hoyle}, {James}, {Krause},
  {Kuehn}, {Lima}, {Maia}, {Marshall}, {Menanteau}, {Miquel}, {Plazas
  Malag{\'o}n}, {Roodman}, {Sanchez}, {Scarpine}, {Schubnell}, {Serrano},
  {Smith}, {Smith}, {Soares-Santos}, {Sobreira}, {Suchyta}, {Swanson}, {Tarle},
  {Vikram}, {Walker}, \& {the DES Collaboration}}]{Liao2020}
{Liao}, W.-T., {Chen}, Y.-C., {Liu}, X., {et~al.} 2020, \mnras, submitted;
  e-print arXiv:2008.12317

\bibitem[{{Lubow} \& {Artymowicz}(1996)}]{Lubow1996}
{Lubow}, S.~H., \& {Artymowicz}, P. 1996, in NATO Advanced Study Institute
  (ASI) Series C, Vol. 477, Evolutionary Processes in Binary Stars, ed.
  R.~A.~M.~J. {Wijers}, M.~B. {Davies}, \& C.~A. {Tout}, 53

\bibitem[{{MacFadyen} \& {Milosavljevi{\'c}}(2008)}]{MacFadyen2008}
{MacFadyen}, A.~I., \& {Milosavljevi{\'c}}, M. 2008, \apj, 672, 83

\bibitem[{{Mart{\'\i}n} {et~al.}(2005){Mart{\'\i}n}, {Magazz{\`u}}, {Delfosse},
  \& {Mathieu}}]{Martin2005}
{Mart{\'\i}n}, E.~L., {Magazz{\`u}}, A., {Delfosse}, X., \& {Mathieu}, R.~D.
  2005, \aap, 429, 939

\bibitem[{{Mathieu} {et~al.}(1991){Mathieu}, {Adams}, \&
  {Latham}}]{Mathieu1991}
{Mathieu}, R.~D., {Adams}, F.~C., \& {Latham}, D.~W. 1991, \aj, 101, 2184

\bibitem[{{Mathieu} {et~al.}(1997){Mathieu}, {Stassun}, {Basri}, {Jensen},
  {Johns-Krull}, {Valenti}, \& {Hartmann}}]{Mathieu1997}
{Mathieu}, R.~D., {Stassun}, K., {Basri}, G., {et~al.} 1997, \aj, 113, 1841

\bibitem[{{Milosavljevi{\'c}} \& {Phinney}(2005)}]{MilosPhinney2005}
{Milosavljevi{\'c}}, M., \& {Phinney}, E.~S. 2005, \apjl, 622, L93

\bibitem[{{Mingarelli}(2019)}]{Mingarelli2019}
{Mingarelli}, C. M.~F. 2019, Nature Astronomy, 3, 8

\bibitem[{{Miranda} {et~al.}(2017){Miranda}, {Mu{\~n}oz}, \&
  {Lai}}]{Miranda2017}
{Miranda}, R., {Mu{\~n}oz}, D.~J., \& {Lai}, D. 2017, \mnras, 466, 1170

\bibitem[{{Mu{\~n}oz} \& {Lai}(2016)}]{Munoz2016}
{Mu{\~n}oz}, D.~J., \& {Lai}, D. 2016, \apj, 827, 43

\bibitem[{{Mu{\~n}oz} {et~al.}(2020){Mu{\~n}oz}, {Lai}, {Kratter}, \& {Mirand
  a}}]{Munoz2020}
{Mu{\~n}oz}, D.~J., {Lai}, D., {Kratter}, K., \& {Mirand a}, R. 2020, \apj,
  889, 114

\bibitem[{{Mu{\~n}oz} {et~al.}(2019){Mu{\~n}oz}, {Miranda}, \&
  {Lai}}]{Munoz2019}
{Mu{\~n}oz}, D.~J., {Miranda}, R., \& {Lai}, D. 2019, \apj, 871, 84

\bibitem[{{Muzerolle} {et~al.}(2019){Muzerolle}, {Flaherty}, {Balog}, {Beck},
  \& {Gutermuth}}]{Muzerolle2019}
{Muzerolle}, J., {Flaherty}, K., {Balog}, Z., {Beck}, T., \& {Gutermuth}, R.
  2019, \apj, 877, 29

\bibitem[{{Oomen} {et~al.}(2018){Oomen}, {Van Winckel}, {Pols}, {Nelemans},
  {Escorza}, {Manick}, {Kamath}, \& {Waelkens}}]{Oomen2018}
{Oomen}, G.-M., {Van Winckel}, H., {Pols}, O., {et~al.} 2018, \aap, 620, A85

\bibitem[{{Peters}(1964)}]{Peters1964}
{Peters}, P.~C. 1964, Physical Review, 136, 1224

\bibitem[{{Pols} {et~al.}(2003){Pols}, {Karakas}, {Lattanzio}, \&
  {Tout}}]{Pols2003}
{Pols}, O.~R., {Karakas}, A.~I., {Lattanzio}, J.~C., \& {Tout}, C.~A. 2003, in
  Astronomical Society of the Pacific Conference Series, Vol. 303, Symbiotic
  Stars Probing Stellar Evolution, ed. R.~L.~M. {Corradi}, J.~{Mikolajewska},
  \& T.~J. {Mahoney}, 290

\bibitem[{{Rafikov}(2016)}]{Rafikov2016}
{Rafikov}, R.~R. 2016, \apj, 830, 8

\bibitem[{{Robson} {et~al.}(2019){Robson}, {Cornish}, \& {Liu}}]{Robson2019}
{Robson}, T., {Cornish}, N.~J., \& {Liu}, C. 2019, Classical and Quantum
  Gravity, 36, 105011

\bibitem[{{Roedig} {et~al.}(2011){Roedig}, {Dotti}, {Sesana}, {Cuadra}, \&
  {Colpi}}]{Roedig2011}
{Roedig}, C., {Dotti}, M., {Sesana}, A., {Cuadra}, J., \& {Colpi}, M. 2011,
  \mnras, 415, 3033

\bibitem[{{Romero-Shaw} {et~al.}(2020){Romero-Shaw}, {Lasky}, {Thrane}, \&
  {Calderon Bustillo}}]{Romero-Shaw+2020}
{Romero-Shaw}, I.~M., {Lasky}, P.~D., {Thrane}, E., \& {Calderon Bustillo}, J.
  2020, arXiv e-prints, arXiv:2009.04771

\bibitem[{{Sesana}(2015)}]{Sesana2015}
{Sesana}, A. 2015, in Gravitational Wave Astrophysics, Vol.~40, 147

\bibitem[{{Tang} {et~al.}(2018){Tang}, {Haiman}, \& {MacFadyen}}]{Tang+2018}
{Tang}, Y., {Haiman}, Z., \& {MacFadyen}, A. 2018, \mnras, 476, 2249

\bibitem[{{Tang} {et~al.}(2017){Tang}, {MacFadyen}, \& {Haiman}}]{Tang2017}
{Tang}, Y., {MacFadyen}, A., \& {Haiman}, Z. 2017, \mnras, 469, 4258

\bibitem[{{Tiede} {et~al.}(2020){Tiede}, {Zrake}, {MacFadyen}, \&
  {Haiman}}]{Tiede2020}
{Tiede}, C., {Zrake}, J., {MacFadyen}, A., \& {Haiman}, Z. 2020, \apj, 900, 43

\bibitem[{Tofflemire {et~al.}(2017)Tofflemire, Mathieu, Ardila, Akeson, Ciardi,
  Johns-Krull, Herczeg, \& Quijano-Vodniza}]{Tofflemire2017a}
Tofflemire, B.~M., Mathieu, R.~D., Ardila, D.~R., {et~al.} 2017, The
  Astrophysical Journal, 835, 8

\bibitem[{{Waters} {et~al.}(1998){Waters}, {Cami}, {de Jong}, {Molster}, {van
  Loon}, {Bouwman}, {de Koter}, {Waelkens}, {Van Winckel}, {Morris}, {Cami},
  {Bouwman}, {de Koter}, {de Jong}, \& {de Graauw}}]{Waters1998}
{Waters}, L.~B.~F.~M., {Cami}, J., {de Jong}, T., {et~al.} 1998, \nat, 391, 868

\bibitem[{{Zrake} {et~al.}(2021){Zrake}, {Tiede}, {MacFadyen}, \&
  {Haiman}}]{Zrake2021}
{Zrake}, J., {Tiede}, C., {MacFadyen}, A., \& {Haiman}, Z. 2021

\end{thebibliography}

\end{document}